\documentclass[12pt]{article}
\usepackage{epsfig,cite,psfrag}

\input paperdef
\newcommand{\psfragtextscale}{0.75}
\psfrag{Mh0}[][][\psfragtextscale]{$m_h [\mathrm{GeV}]$}
\psfrag{MA0}[][][\psfragtextscale]{$M_A [\mathrm{GeV}]$}
\psfrag{MHp}[][][\psfragtextscale]{$M_{H^\pm} [\mathrm{GeV}]$}
\psfrag{MH}[][][\psfragtextscale]{$m_{h_i} [\mathrm{GeV}]$}
\psfrag{TB}[][][\psfragtextscale]{$\tan \beta$}
\psfrag{Mh1}[][][\psfragtextscale]{$m_{h_1} [\mathrm{GeV}]$}
\psfrag{dMh1}[][][\psfragtextscale]{$\Delta m_{h_1} [\mathrm{GeV}]$}
\psfrag{dMH2H3}[][][\psfragtextscale]{$\Delta_{32} [\mathrm{GeV}]$}
\psfrag{APart}[][][\psfragtextscale]{$(\cp{\rm -odd~part})^2$}
\psfrag{argMUE}[][][\psfragtextscale]{$\phi_\mu$}
\psfrag{argM2}[][][\psfragtextscale]{$\phi_{M_2}$}
\psfrag{argAt}[][][\psfragtextscale]{$\phi_{\At}$}
\psfrag{argXt}[][][\psfragtextscale]{$\phi_{\Xt}$}
\psfrag{ghVV}[][][\psfragtextscale]{$g_{hVV}$}
\psfrag{Uij}[][][\psfragtextscale]{$u_{ij}$}
\psfrag{5x-1}[][][\psfragtextscale]{0.5}
\psfrag{1x-1}[][][\psfragtextscale]{$10^{-1}$}
\psfrag{5x-2}[][][\psfragtextscale]{$5 \, 10^{-2}$}
\psfrag{1x-2}[][][\psfragtextscale]{$10^{-2}$}
\psfrag{1x0}[][][\psfragtextscale]{1}
\psfrag{5x0}[][][\psfragtextscale]{5}
\psfrag{1x1}[][][\psfragtextscale]{10}
\psfrag{5x1}[][][\psfragtextscale]{50}
\psfrag{0}[][][\psfragtextscale]{0}
\psfrag{0.1}[][][\psfragtextscale]{0.1}
\psfrag{0.2}[][][\psfragtextscale]{0.2}
\psfrag{0.3}[][][\psfragtextscale]{0.3}
\psfrag{0.4}[][][\psfragtextscale]{0.4}
\psfrag{0.5}[][][\psfragtextscale]{0.5}
\psfrag{0.6}[][][\psfragtextscale]{0.6}
\psfrag{0.7}[][][\psfragtextscale]{0.7}
\psfrag{0.8}[][][\psfragtextscale]{0.8}
\psfrag{0.9}[][][\psfragtextscale]{0.9}
\psfrag{1}[][][\psfragtextscale]{1}
\psfrag{1.}[][][\psfragtextscale]{1}
\psfrag{1.5}[][][\psfragtextscale]{1.5}
\psfrag{2}[][][\psfragtextscale]{2}
\psfrag{3}[][][\psfragtextscale]{3}
\psfrag{4}[][][\psfragtextscale]{4}
\psfrag{5}[][][\psfragtextscale]{5}
\psfrag{10}[][][\psfragtextscale]{10}
\psfrag{20}[][][\psfragtextscale]{20}
\psfrag{30}[][][\psfragtextscale]{30}
\psfrag{40}[][][\psfragtextscale]{40}
\psfrag{50}[][][\psfragtextscale]{50}
\psfrag{60}[][][\psfragtextscale]{60}
\psfrag{70}[][][\psfragtextscale]{70}
\psfrag{80}[][][\psfragtextscale]{80}
\psfrag{90}[][][\psfragtextscale]{90}
\psfrag{100}[][][\psfragtextscale]{100}
\psfrag{110}[][][\psfragtextscale]{110}
\psfrag{120}[][][\psfragtextscale]{120}
\psfrag{130}[][][\psfragtextscale]{130}
\psfrag{140}[][][\psfragtextscale]{140}
\psfrag{150}[][][\psfragtextscale]{150}
\psfrag{160}[][][\psfragtextscale]{160}
\psfrag{170}[][][\psfragtextscale]{170}
\psfrag{180}[][][\psfragtextscale]{180}
\psfrag{190}[][][\psfragtextscale]{190}
\psfrag{200}[][][\psfragtextscale]{200}
\psfrag{250}[][][\psfragtextscale]{250}
\psfrag{300}[][][\psfragtextscale]{300}
\psfrag{400}[][][\psfragtextscale]{400}
\psfrag{500}[][][\psfragtextscale]{500}
\psfrag{600}[][][\psfragtextscale]{600}
\psfrag{700}[][][\psfragtextscale]{700}
\psfrag{800}[][][\psfragtextscale]{800}
\psfrag{900}[][][\psfragtextscale]{900}
\psfrag{1000}[][][\psfragtextscale]{1000}
\psfrag{1500}[][][\psfragtextscale]{1500}
\psfrag{-0.5}[][][\psfragtextscale]{-0.5}
\psfrag{-1}[][][\psfragtextscale]{-1}
\psfrag{-1.}[][][\psfragtextscale]{-1}
\psfrag{-1.5}[][][\psfragtextscale]{-1.5}
\psfrag{-2}[][][\psfragtextscale]{-2}
\psfrag{-3}[][][\psfragtextscale]{-3}
\psfrag{-4}[][][\psfragtextscale]{-4}
\psfrag{-5}[][][\psfragtextscale]{-5}
\psfrag{-10}[][][\psfragtextscale]{-10}
\psfrag{-20}[][][\psfragtextscale]{-20}
\psfrag{-30}[][][\psfragtextscale]{-30}
\psfrag{-40}[][][\psfragtextscale]{-40}
\psfrag{-50}[][][\psfragtextscale]{-50}
\psfrag{-60}[][][\psfragtextscale]{-60}
\psfrag{-70}[][][\psfragtextscale]{-70}
\psfrag{-80}[][][\psfragtextscale]{-80}
\psfrag{-90}[][][\psfragtextscale]{-90}
\psfrag{-100}[][][\psfragtextscale]{-100}
\psfrag{-110}[][][\psfragtextscale]{-110}
\psfrag{-120}[][][\psfragtextscale]{-120}
\psfrag{-130}[][][\psfragtextscale]{-130}
\psfrag{-140}[][][\psfragtextscale]{-140}
\psfrag{-150}[][][\psfragtextscale]{-150}
\psfrag{-160}[][][\psfragtextscale]{-160}
\psfrag{-170}[][][\psfragtextscale]{-170}
\psfrag{-180}[][][\psfragtextscale]{-180}
\psfrag{-190}[][][\psfragtextscale]{-190}
\psfrag{-200}[][][\psfragtextscale]{-200}
\psfrag{-300}[][][\psfragtextscale]{-300}
\psfrag{-400}[][][\psfragtextscale]{-400}
\psfrag{-500}[][][\psfragtextscale]{-500}
\psfrag{-600}[][][\psfragtextscale]{-600}
\psfrag{-700}[][][\psfragtextscale]{-700}
\psfrag{-800}[][][\psfragtextscale]{-800}
\psfrag{-900}[][][\psfragtextscale]{-900}
\psfrag{-1000}[][][\psfragtextscale]{-1000}
\psfrag{-1500}[][][\psfragtextscale]{-1500}

\begin{document}
\thispagestyle{empty}

\def\thefootnote{\fnsymbol{footnote}}

\begin{flushright}
DCPT/02/128\\ 
IPPP/02/64\\
KA-TP--22--2002\\
LMU 16/02\\
MPI-PhT/2002-74\\
hep-ph/0212037
\end{flushright}

\vspace{.5cm}

\begin{center}

{\large\sc {\bf The Higgs Boson Masses of the Complex MSSM:}}

\vspace{0.4cm}

{\large\sc {\bf A Complete One-Loop Calculation}}%
\footnote{Talk given by S.~Heinemeyer at SUSY02, June 2002, DESY, Germany}

\vspace{1cm}

{\sc 
M.~Frank$^{1}$%
\footnote{email: Markus.Frank@physik.uni-karlsruhe.de}%
, S.~Heinemeyer$^{2}$%
\footnote{email: Sven.Heinemeyer@physik.uni-muenchen.de}%
, W.~Hollik$^{3}$%
\footnote{email: hollik@mppmu.mpg.de}%
~and G.~Weiglein$^{4}$%
\footnote{email: Georg.Weiglein@durham.ac.uk}
}

\vspace*{1cm}

{\sl
$^1$ Institut f\"ur Theoretische Physik, Universit\"at Karlsruhe, \\
D--76128 Karlsruhe, Germany

\vspace*{0.4cm}

$^2$Institut f\"ur theoretische Elementarteilchenphysik,
LMU M\"unchen, Theresienstr.\ 37, D--80333 M\"unchen, Germany

\vspace*{0.4cm}

$^3$Max-Planck-Institut f\"ur Physik (Werner-Heisenberg-Institut),
F\"ohringer Ring 6, \\
D--80805 M\"unchen, Germany

\vspace*{0.4cm}

$^4$Institute for Particle Physics Phenomenology, University of Durham,\\
Durham DH1~3LE, UK
}

\end{center}

\vspace*{0.2cm}

\begin{abstract}
  In the Minimal Supersymmetric Standard Model with complex parameters\\
  (cMSSM) we perform a complete \onel\ calculation for the Higgs boson
  masses (including the momentum dependence)
  and the mixing angles. These corrections are obtained 
  in the Feynman-diagrammatic approach using the on-shell
  renormalization scheme. 
  The impact of the newly evaluated corrections is analyzed
  numerically. The full \onel\ result, supplemented by the 
  leading \twol\ contributions taken over from the real MSSM 
  are implemented into the public Fortran code \fh2.0.
\end{abstract}

\def\thefootnote{\arabic{footnote}}
\setcounter{page}{0}
\setcounter{footnote}{0}

\newpage


\section{Introduction}

The search for the lightest Higgs boson is a crucial test of 
Supersymmetry (SUSY) which can be performed with the present and the
next generation of accelerators. The prediction of a relatively
light Higgs boson is common to all supersymmetric models whose
couplings remain in the perturbative regime up to a very high energy
scale~\cite{susylighthiggs}.
A precise prediction for the mass of the lightest Higgs boson and its
couplings to other particles in terms
of the relevant SUSY parameters is necessary in order to determine the
discovery and exclusion potential of LEP2 and the upgraded Tevatron, and
for physics at the LHC and future linear colliders, where eventually a
high-precision measurement of the properties of the Higgs boson might
be possible~\cite{lc}. 

The case of the Higgs sector in the $\cp$-conserving MSSM (rMSSM) has been
tackled up to the \twol\ level by different methods such as the
Effective Potential (EP) method~\cite{mhiggsEP,mhiggsEP2,mhiggsEP3,mhiggsEP2L},
the renormalization 
group (RG) improved \onel\ EP approach~\cite{mhiggsRG} and the
Feynman-diagrammatic (FD) method using the on-shell renormalization
scheme~\cite{mhiggsletter,mhiggslong}. 
This method has provided the only complete calculation at
the \onel\ level including momentum dependence~\cite{mhiggsFD1l} and
furthermore the dominant logarithmic and non-logarithmic 
corrections at the \twol\ level~\cite{mhiggslong,bse}.
The application of different
methods lead to thorough comparisons between the different approaches.

In the case of the MSSM with complex parameters (cMSSM) the higher
order corrections have been performed, after the first more
general investigations~\cite{mhiggsCPXgen}, in the EP
approach and in the RG improved \onel\ EP 
method~\cite{mhiggsCPXEP1,mhiggsCPXEP2,schwachnath,mhiggsCPXRG1,mhiggsCPXRG2}.
In the context of the FD 
approach, so far an exploratory study, including a calculation of the
leading terms has been performed~\cite{mhcMSSM}. A full \onel\
calculation, including momentum dependence, as well as an evaluation
of the leading \twol\ corrections have been missing so far.

This paper provides the next step into this direction: 
We present the full \onel\ calculation for the Higgs boson masses and
mixing angles in the cMSSM. Concerning the Higgs boson masses, 
the full momentum dependence has been taken into account. 
In this paper we present briefly the calculation and discuss the
effects on the Higgs boson masses that originate from 
the new terms that extend the available calculations obtained using
the EP and RG approach. More details about the evaluation as well as a
more detailed numerical analysis can be found in \citeres{mhcMSSMlong,MFphd}.  
All results are incorporated into a public Fortran code, \fh2.0. 

The rest of the paper is organized as follows. In Section~2 we review
the Higgs sector and the scalar quark sector of the cMSSM, providing
the relations of physical and
unphysical parameters, the masses and the mixing
angles. Section~3 contains the numerical analysis. We conclude with
Section~4.


\section{Calculational framework}

\subsection{The tree-level structure of the cMSSM Higgs sector}

The cMSSM Higgs potential reads~\cite{hhg}:
\BEA
\label{Higgspot}
V &=& m_1^2 \cHe\bar{\cHe} + m_2^2 \cHz\bar{\cHz} - (m_{12}^2 \epsilon_{ab}
      \cHe^a\cHz^b + \hc)  \nonumber \\
   && \mbox{} + \frac{g'^2 + g^2}{8}\, (\cHe\bar{\cHe} - \cHz\bar{\cHz})^2
      +\frac{g^2}{2}\, |\cHe\bar{\cHz}|^2,
\EEA
where $m_1^2, m_2^2, m_{12}^2$ are soft SUSY-breaking terms ($m_{12}^2$
can be complex), $g, g'$
are the $SU(2)$ and $U(1)$ gauge couplings, and $\epsilon_{12} = -1$.
The doublet fields $\cHe$ and $\cHz$ are decomposed  in the following way:
\BEA
\cHe &=& \VL H_1^1 \\ H_1^2 \VR = \VL v_1 + (\phi_1^{0} + i\chi_1^{0})
                                 /\sqrt2 \\ \phi_1^- \VR ,\non \\
\cHz &=& \VL H_2^1 \\ H_2^2 \VR =  e^{i\xi} \VL \phi_2^+ \\ v_2 + (\phi_2^0 
                                              + i\chi_2^0)/\sqrt2 \VR.
\label{eq:hidoubl}
\EEA
$\xi$ is a phase between the two Higgs doublets. In the
Higgs potential it appears only in the combination 
$\xi' = \xi + \arg(m_{12}^2)$. From the
unphysical parameters in \refeq{Higgspot} the transition to the
physical parameters (including the tadpoles) is performed by the
substitution (see \citeres{mhcMSSMlong,MFphd} for details):
\BE
v_1,\; v_2,\; g_1,\; g_2,\; m_1^2,\; m_2^2,\; |m_{12}^2|,\; \xi'
\; \to \;
e,\; \sw,\; \MZ,\; \tb,\; \MHp^2,\; T_h,\; T_H,\; T_A ~.
\label{subst} 
\EE
$e$ is the electric charge, $\sw^2 = 1 - \cw^2$ with 
$\cw \equiv \MW/\MZ$, where $\MW$ and $\MZ$ are the masses of the
$W$~and the $Z$~boson, respectively. Furthermore, 
$\tb$ is the ratio of the two vacuum expectation values, $\tb =
v_2/v_1$ ($\sbe \equiv \Sb, \cbe \equiv \Cb$), $\MHp$ is the mass of
the charged Higgs boson, and $T_x, x = h, H, A$ denote the tadpoles of
the fields $h$, $H$ and $A$, see below. 

At the tree-level, the tadpoles have to
be zero. In the case of $T_A$ this can only be fulfilled if 
$\xi' = 0$. Thus $\cp$-violation is absent at the tree-level.
The bilinear part of the Higgs potential has to be diagonalized to
obtain the mass eigenstates. In the $\cp$-even sector this is done
with the help of the angle $\al$, and results in the two neutral
$\cp$-even Higgs bosons $h$ and $H$. In the $\cp$-odd sector the
diagonalization can be performed with the angle $\be$, leading to the
$\cp$-odd $A$~boson and the neutral Goldstone boson~$G$. The charged
Higgs sector is also diagonalized with the angle $\be$, resulting in
the charged Higgs bosons~$H^\pm$ and the Goldstone bosons~$G^\pm$.
One arrives at the following masses at tree-level:
\BEA
H &:& \mH^2 = \edz \KKL \MHpq^2 + \MZ^2 
                +\sqrt{(\MHpq^2 + \MZ^2)^2 
                       - 4 \MZ^2\MHpq^2 c_{2\be}^2}~\KKR \non \\
h &:& \mh^2 = \edz \KKL \MHpq^2 + \MZ^2 
                -\sqrt{(\MHpq^2 + \MZ^2)^2 
                       - 4 \MZ^2\MHpq^2 c_{2\be}^2}~\KKR \non \\
A &:& \mA^2 = \MHp^2 - \MW^2 \quad (\equiv \MHpq^2) \non \\
G &:& m_G^2 = \MZ^2 \non \\
H^\pm &:& \MHp^2 \non \\ 
G^\pm &:& m_{G^\pm}^2 = \MW^2
\EEA
$\MHp$ and $\tb$ are chosen as input parameters.
The entries for the Goldstone bosons $G$ and $G^\pm$ are to be
understood in the Feynman gauge.
Since there  is no $\cp$-violation in the cMSSM Higgs sector at
tree-level, there is no mixing between $h$ and $H$ and the fields $A$ 
and~$G$.


\subsection{The fermion and gaugino sectors of the cMSSM}
\label{cMSSMsectors}

Possibly $\cp$-violating parameters occur in all other SUSY sectors of the
cMSSM. Most important for Higgs boson phenomenology is the scalar
quark sector. 
The mass matrix of two squarks of the same flavor, 
$\sql$ and $\sqr$, is given by
\BE
M_{\sq} = \ML M_L^2 + \mq^2 & \mq \; \Xq^* \\
            \mq \; \Xq    & M_R^2 + \mq^2 \MR
\label{squarkmassmatrix}
\EE
with 
\BEA
M_L^2 &=& M_{\tilde Q}^2 + \MZ^2 \CZb (I_3^q - Q_q \sw^2) \non \\
M_R^2 &=& M_{\tilde Q'}^2 + \MZ^2 \CZb Q_q \sw^2 \\ \non 
\Xq &=& A_q - \mu^* \{\CTb, \tb\} ,
\label{squarksoftSUSYbreaking}
\EEA
where $\{\CTb, \tb\}$ applies for $\{ {\rm up, down} \}$-type 
squarks, respectively. $A_q$ and $\mu$ can be complex.
In an isodoublet the $SU(2)$ symmetry enforces that $M_{\tilde Q}$ has
to be chosen equal for both squark types. The $M_{\tilde Q'}$ on the
other hand can be chosen independently for every squark type.
In the scalar quark sector of the cMSSM $N_q + 1$ phases are
present, one for each $A_q$ and one 
for $\mu$.
The squark mass eigenstates are obtained by diagonalizing the mass
matrix and are given by
\BE
m_{\tilde q_{1,2}}^2 = \mq^2 
  + \edz \KKL M_L^2 + M_R^2 
           \mp \sqrt{( M_L^2 - M_R^2)^2 + 4 \mq^2 |\Xq|^2}~\KKR .
\EE
The masses are independent of the phase of $\Xq$. However, 
the phase of $A_q$ affects the mass eigenvalues since it changes the
absolute value of $\Xq$.

The other possibly complex parameters are $M_1$ and $M_2$, which are
the soft SUSY-breaking parameters in the gaugino sector, and $\mgl$,
the gluino mass. These parameters are less important for the Higgs
sector phenomenology; $\mgl$ enters the predictions only at the \twol\
level.


\subsection{The neutral Higgs boson sector at the \onel\ level}
\label{subsec:Higgs1L}

The inverse neutral Higgs boson propagator matrix in the cMSSM at the
\onel\ level is given by
\BE
(\De_{\rm Higgs})^{-1} = -i
\MLd q^2 - \mh^2 + \hSi_{hh}(q^2) & \hSi_{hH}(q^2) & \hSi_{hA}(q^2) \\
     \hSi_{Hh}(q^2) & q^2 - \mH^2 + \hSi_{HH}(q^2) & \hSi_{HA}(q^2) \\
     \hSi_{Ah}(q^2) & \hSi_{AH}(q^2) & q^2 - \mA^2 + \hSi_{AA}(q^2) \MR~.
\EE
$\hSi$ denotes the renormalized Higgs boson self-energies (at the
\onel\ level). 
$\cp$-violation occurs, i.e.\ mixing between the $\cp$-even Higgs
bosons $h$, $H$, and the $\cp$-odd Higgs boson $A$ occurs if the
self-energies $\hSi_{AH} = \hSi_{HA}$ and/or $\hSi_{Ah} = \hSi_{hA}$
are non-zero. This can happen if the complex parameters in the cMSSM
possess an imaginary part. Also the pure $\cp$-even self-energies,
$\hSi_{hh}$, $\hSi_{HH}$ and $\hSi_{hH}$, and the pure $\cp$-odd
self-energy, $\hSi_{AA}$, are numerically affected if
some cMSSM parameters are complex. Details about the calculation of
the renormalized Higgs boson self-energies and the on-shell
renormalization procedure can be found in
\citeres{mhcMSSM,mhcMSSMlong,MFphd}. 

The poles of the Higgs boson propagator matrix are the squares of the
pole masses of the three neutral Higgs bosons including higher-order
corrections. The masses are denotes by
$\mhe$, $\mhz$, $\mhd$ with $\mhe \le \mhz \le \mhd$.

Besides the Higgs boson masses, also the effects of the higher-order
corrected self-energies on the Higgs boson couplings to SM gauge
bosons and fermions can be evaluated. In the limit of $q^2 = 0$, i.e.\
$\hSi(q^2) \to \hSi(0)$, the transition from the tree-level states 
$h, H, A$ to the mass eigenstates at higher orders, $h_1, h_2, h_3$
can be described with the rotation matrix $U$,
\BE
\VL h_1 \\ h_2 \\ h_3 \VR =
\MLd u_{11} & u_{12} & u_{13} \\
     u_{21} & u_{22} & u_{23} \\
     u_{31} & u_{32} & u_{33} \MR
\VL h \\ H \\ A \VR \quad 
\equiv U \VL h \\ H \\ A \VR~.
\EE
$U$ includes the dominant corrections (coming from Higgs boson
propagators) into the effective couplings~\cite{hff}. The
explicit formulas are given in \citeres{mhiggsCPXRG1,mhcMSSM}. 

\bigskip
The Higgs boson self-energies have been evaluated by taking into account
all sectors of the cMSSM, including possibly complex parameters and
the full momentum dependence. The evaluation has been done with the
help of the programs \fa~\cite{feynarts} (using the MSSM model
file~\cite{modelfile}) and \fc~\cite{formcalc}. 
The results, supplemented by the leading \twol\ contributions taken
over from the rMSSM~\cite{mhiggsletter,mhiggslong,mhiggsEP2}, have
been transformed into the Fortran code \fh2.0~\cite{feynhiggs}.
The code evaluates the Higgs boson masses, the mixing matrix and the
corresponding corrections to the couplings of the Higgs bosons to SM
gauge bosons and fermions.
\fh2.0 is available at {\tt www.feynhiggs.de} .


\section{Numerical examples and discussion}

In this section numerical examples are presented and discussed.
They are meant
to illustrate on the one hand the possible effects of complex phases
in the MSSM and on the other hand the effects of the newly evaluated
terms. For a more detailed phenomenological analysis
constraints on $\cp$-violating parameters from experimental bounds,
e.g.\ on electric 
dipole moments (EDMs), have to be taken into account~\cite{pdg}.
However, in our analysis below we only take non-zero phases for 
$\At = \Ab$ and $M_2$, which are not severely restricted from EDM
bounds. 

The numerical analysis given below has been performed in the
``CPX''~scenario~\cite{cpx}, where the parameters are fixed to
\BEA
&& \msusy = 500 \gev, \; |\At| = |\Ab| = 1000 \gev, \; 
   \mu = 2000 \gev, \; M_2 = 500 \gev, \non \\
&& \MHp = 150 \gev, \; \tb = 5 ~,
\label{cpx}
\EEA
if not indicated differently. The phases are always defined
explicitly. However, our analysis is confined to 
$\phi_\mu = 0$, since this is the most restricted phase, see e.g.\
\citere{plehnix} and 
references therein. The CPX~scenario has been defined in order
to maximize the effects of complex phases. It should be kept in mind
that while the relatively high value of $|\mu|$ is not realized in
minimal models like mSUGRA, mGMSB or mAMSB, it can lead to large
effects especially in the $b/\Sbot$~sector. 

In our numerical analysis we concentrate on the Higgs boson masses
derived from the \onel\ corrections
only. This is sufficient to show the effects of the
newly evaluated corrections. 
A more detailed numerical analysis, including also the effects on the
rotation matrix and correspondingly to the Higgs boson couplings to SM
gauge bosons and fermions can be found in \citeres{mhcMSSMlong,MFphd}.

\begin{figure}[htb!]
\vspace{2em}
\begin{center}
\includegraphics{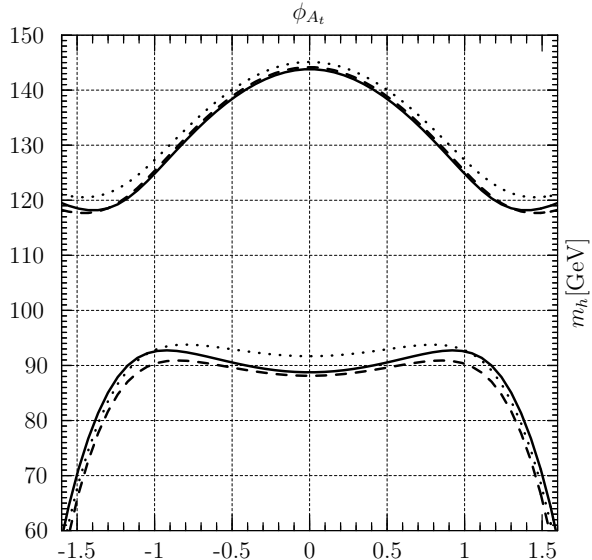}
\caption{
$\mhe$ and $\mhz$ are shown as a function of $\phi_{\At}$ for the
parameters of \refeq{cpx}. The dotted line shows the
result for the (s)fermion sector in the $q^2 = 0$
approximation, the dashed line shows the full cMSSM with $q^2 = 0$,
and the full line also includes the effects of the non-vanishing momentum.
}
\label{fig:mh1mh2_PhiAt}
\end{center}
\end{figure}

In \reffi{fig:mh1mh2_PhiAt} we show the two lighter masses, $\mhe$ and
$\mhz$, for the parameters in \refeq{cpx} as a function of
$\phi_{\At}$. The dotted line shows the
result for the (s)fermion sector%
\footnote{
With the (s)fermion sector we denote here the $t/\Stop/b/\Sbot$~sector.
The effects of the other fermions and sfermions are numerically small
and stay below $1 \gev$~\cite{mhcMSSMlong,MFphd}.
}%
~in the $q^2 = 0$
approximation, the dashed line shows the full cMSSM with $q^2 = 0$,
and the full line also includes the momentum contributions.
The effect of the subleading \onel\ contributions, i.e.\ the ones
beyond the (s)fermion sector, can give rise to changes in $\mhe$
and $\mhz$ of \order{4 \gev}. The further inclusion of the momentum
dependence can result in a shift of $\mhe$ of \order{2 \gev} for
large values of $\phi_{\At}$. 
%

\begin{figure}[htb!]
\vspace{1em}
\begin{center}
\includegraphics{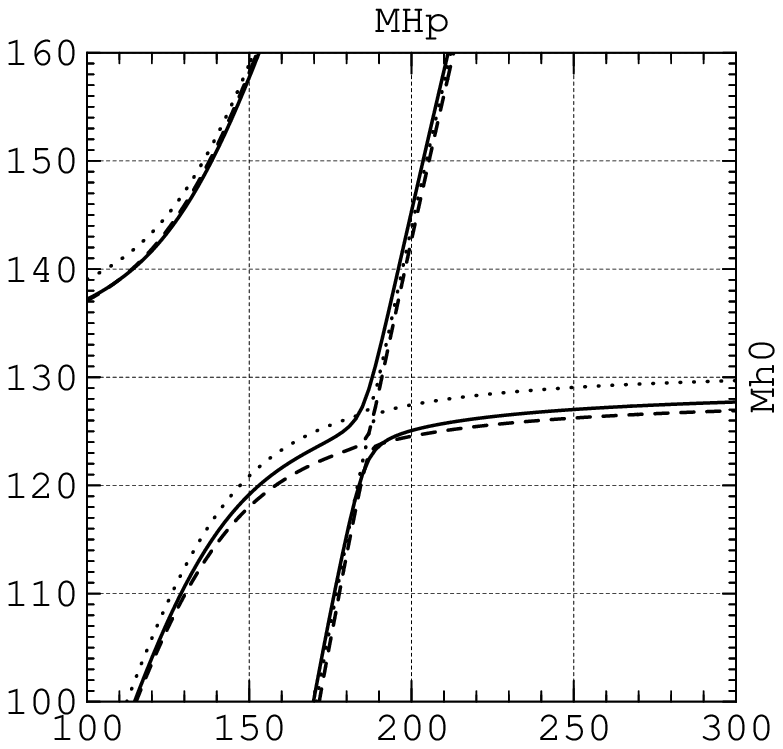}
\caption{
$\mhe$, $\mhz$ and $\mhd$ are shown as a function of $\MHp$ for the
parameters of \refeq{cpx} and $\phi_{\At} = \pi/2$. The line styles are as
in \reffi{fig:mh1mh2_PhiAt}.
}
\label{fig:mh1mh2mh3_MHp}
\end{center}
\end{figure}

\begin{figure}[htb!]
\vspace{1em}
\begin{center}
\includegraphics{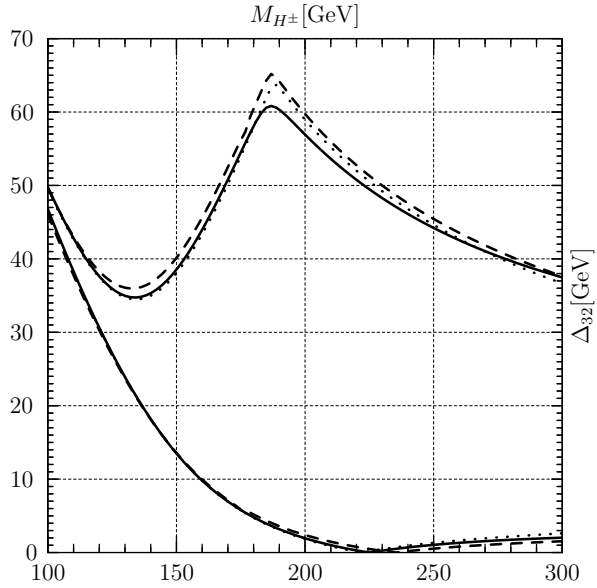}
\caption{
$\De_{32} = \mhd - \mhz$ is shown as a function of $\MHp$ for the
parameters of \refeq{cpx}. The line styles are as
in \reffi{fig:mh1mh2_PhiAt}. The upper and lower curves correspond to
$\phi_{\At} = \pi/2$ and $\phi_{\At} = 0$.
}
\label{fig:Deltamh3mh2_MHp}
\end{center}
\end{figure}

All neutral Higgs boson masses, $\mhe$, $\mhz$, $\mhd$, are shown as a
function of $\MHp$ 
in \reffi{fig:mh1mh2mh3_MHp} for the parameters of \refeq{cpx}. It
becomes apparent that especially when two mass eigenvalues are close to
each other, the newly evaluated terms and the momentum dependent
contribution can change the results by several GeV. Especially the
inclusion of the momentum dependence results here in a larger mass gap
of the two lightest masses compared to the result where the momentum
dependence has been neglected.

Of particular interest is the mass gap between the two heavier neutral
Higgs boson masses. For values of $\MHp \gsim 200 \gev$, $\mhz$ and
$\mhd$ can be very close to each other, which makes their experimental
resolution at a collider experiment difficult. In
\reffi{fig:Deltamh3mh2_MHp} we show $\De_{32} = \mhd - \mhz$ as a
function of $\MHp$ in the rMSSM, i.e.\ for $\phi_{\At} = 0$ (lower set
of curves) and for $\phi_{\At} = \pi/2$ (upper set). The different
line styles are as in \reffi{fig:mh1mh2_PhiAt}. In the chosen scenario
$\De_{32}$ is always larger in the case of $\phi_{\At} = \pi/2$ as
compared to the case where $\phi_{\At} = 0$. The induced difference in
$\De_{32}$ can be 
larger than $50 \gev$. The impact of the non-(s)fermionic terms can be
of \order{5 \gev}. It should be kept in mind that changing the phase
of $\At$ also effects the absolute value of $\Xt$. In fact, in
\citeres{mhcMSSMlong,MFphd} it is 
demonstrated that every mass gap that can appear in the cMSSM can (for
another choice of parameters) also be accommodated in the rMSSM.

\begin{figure}[htb!]
\begin{center}
\includegraphics{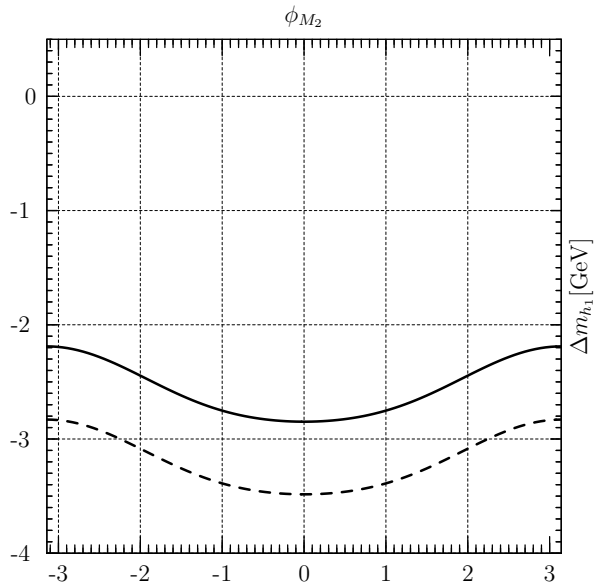}
\caption{
The effects on $\mhe$ of the different sectors normalized to the
(s)fermion sector evaluation only (i.e. the
(s)fermion sector contribution subtracted) is shown as a function of
$\phi_{M_2}$ for $\phi_\mu = \phi_{\At} = 0$ and the other parameters
as in \refeq{cpx}. The dashed (full) line shows the result for the
full cMSSM with $q^2 = 0\; (\neq 0)$. 
}
\label{fig:Deltamh1_phiM2}
\end{center}
\end{figure}

Finally, in \reffi{fig:Deltamh1_phiM2} we show the effect of
$\phi_{M_2}$ on the lightest Higgs boson mass in the scenario of
\refeq{cpx}. The dashed (full) line
shows the difference of the full MSSM with $q^2 = 0 \; (\neq 0)$ to the 
(s)fermion approximation with $q^2 = 0$. The effect of the
%
%
non-(s)fermionic contribution are in this case of \order{3 \gev}, and
the momentum dependence induces a shift of \order{1 \gev}. The effect
of $\phi_{M_2}$ is $\lsim 1 \gev$. Thus the phase of $M_2$ has a
much smaller impact than the phases of the trilinear couplings. In a
more detailed analysis of the cMSSM parameter space (see
\citeres{mhcMSSMlong,MFphd}) it has been found that large effects of
the phases of the gaugino mass parameters on $\mhe, \mhz, \mhd$ or on
the mixing of $\cp$-even and $\cp$-odd states can only occur if 
two mass eigenvalues are very close to each other. Thus, the effect observed
in \citere{schwachnath} that the mixing of the two heavier Higgs
states depends strongly on the phase of $M_2$ or $M_1$ happens only in
a very small part of the cMSSM parameter space.


\section{Conclusions}

We have presented a complete \onel\ calculation of the Higgs boson
masses in the MSSM with complex parameters. The calculation has been
performed in the Feynman-diagrammatic approach, using the on-shell
renormalization scheme. Besides the full spectrum of cMSSM particles
also the momentum dependence has explicitly been included. 

In the numerical analysis we have investigated the effects of the
non-(s)fermionic sectors as well as the effects of the momentum
dependence on the 
three neutral Higgs boson mass eigenvalues, $\mhe$, $\mhz$ and
$\mhd$. The analysis has been performed for a set 
of cMSSM parameters that maximizes the effects of the $\cp$-violating
phases. In this case we find that the corrections of the
non-(s)fermionic sector can be of \order{5 \gev}, while the momentum
dependence induces a shift of up to $2 \gev$. These effects are
more pronounced if two of the mass eigenvalues are close to each other. 
The observed effects on $\mhe$, $\mhz$ and $\mhd$ of the newly included
corrections are thus of the order of several GeV and have to be taken
into account in phenomenological analyses of the cMSSM Higgs boson
sector in view of the prospective experimental precision at the next
generation of colliders. The phase of the gaugino mass parameters,
$\phi_{M_2}$ and 
$\phi_{M_1}$, are found to have only a relatively small effect, except
in a very small regions of the parameter space, where two of the mass
eigenvalues are very close to each other. 

The results of the full \onel\ calculation, supplemented with the
dominant and subdominant corrections taken over from the rMSSM, have
been implemented into the Fortran program \fh2.0. The code can be
obtained at {\tt www.feynhiggs.de} .


\subsection*{Acknowledgments}
We thank C.~Schappacher for helpful discussions. This
work was partially supported by the European Community's
Human Potential Programme under contract 
HPRN-CT-2000-00149 (Collider Physics).
%




\begin{thebibliography}{99} 

\bibitem{susylighthiggs} G.~Kane, C.~Kolda and J.~Wells,
                         {\em Phys. Rev. Lett.} {\bf 70} (1993) 2686,
                         hep-ph/9210242;\\
                         J.~Espinosa and M.~Quir\'os,
                         {\em Phys. Rev. Lett.} {\bf 81} (1998) 516,
                         hep-ph/9804235.

\bibitem{lc} J.~Aguilar-Saavedra et al.,
             TESLA TDR Part~3: 
             ``Physics at an $e^+e^-$ Linear Collider,''
             hep-ph/0106315,
             see: {\tt tesla.desy.de/tdr} ;\\
             T.~Abe {\it et al.}  
             [American Linear Collider Working Group Collaboration],
             ``Linear collider physics resource book for Snowmass 2001,''
             hep-ex/0106056;\\
             K.~Abe et al. 
             [ACFA LC Working Group Collaboration],
             hep-ph/0109166.

\bibitem{mhiggsEP} R.~Hempfling and A.~Hoang, 
                   {\em Phys. Lett.} {\bf B 331} (1994) 99, 
                   hep-ph/9401219;\\
                   R.~Zhang, 
                   {\em Phys. Lett.} {\bf B 447} (1999) 89, 
                   hep-ph/9808299;\\
                   J.~Espinosa and R.~Zhang,
                   {\em Nucl. Phys.} {\bf B 586} (2000) 3,
                   hep-ph/0003246.

\bibitem{mhiggsEP2} G.~Degrassi, P.~Slavich, and F.~Zwirner,
                    {\em Nucl. Phys.} {\bf B 611} (2001) 403,
                    hep-ph/0105096;\\
                    A.~Brignole, G.~Degrassi, P.~Slavich, and F.~Zwirner,
                    {\em Nucl. Phys.} {\bf B 631} (2002) 195,
                    hep-ph/0112177.

\bibitem{mhiggsEP3} A.~Brignole, G.~Degrassi, P.~Slavich, and F.~Zwirner,
                    {\em Nucl. Phys.} {\bf B 643} (2002) 79, 
                    hep-ph/0206101.

\bibitem{mhiggsEP2L} S.~Martin,
                     {\em Phys. Rev.} {\bf D 65} (2002) 116003,
                     hep-ph/0111209;
                     {\em Phys. Rev.} {\bf D 66} (2002) 096001,
                     hep-ph/0206136;
                     hep-ph/0211366.

\bibitem{mhiggsRG} M.~Carena, M.~Quir\'os and C.~Wagner, 
                   {\em Nucl. Phys.} {\bf B 461} (1996) 407, 
                   hep-ph/9508343;\\
                   H.~Haber, R.~Hempfling and A.~Hoang, 
                   {\em Z. Phys.} {\bf C 75} (1997) 539, 
                   hep-ph/9609331.

\bibitem{mhiggsletter} S.~Heinemeyer, W.~Hollik and G.~Weiglein, 
                       {\em Phys. Rev.} {\bf D 58} (1998) 091701, 
                       hep-ph/9803277; 
                       {\em Phys. Lett.} {\bf B 440} (1998) 296, 
                       hep-ph/9807423;
                       {\em Phys. Lett.} {\bf B 455} (1999) 179,
                       hep-ph/9903404.

\bibitem{mhiggslong} S.~Heinemeyer, W.~Hollik and G.~Weiglein, 
                     {\em Eur. Phys. Jour.} {\bf C 9} (1999) 343, 
                     hep-ph/9812472.

\bibitem{mhiggsFD1l} A.~Dabelstein, 
                     {\em Nucl. Phys.} {\bf B 456} (1995) 25, 
                     hep-ph/9503443;
                     {\em Z. Phys.} {\bf C 67} (1995) 495, 
                     hep-ph/9409375;\\
                     P. Chankowski, S. Pokorski, J. Rosiek, {\em Nucl. Phys.}
                     {\bf B 423} (1994) 437,
                     hep-ph/9303309.

\bibitem{bse} M.~Carena, H.~Haber, S.~Heinemeyer, W.~Hollik, C.~Wagner
              and G.~Weiglein, 
              {\em Nucl. Phys.} {\bf B 580} (2000) 29, 
              hep-ph/0001002.

\bibitem{mhiggsCPXgen} A.~Pilaftsis,
                       {\em Phys. Rev.} {\bf D 58} (1998) 096010,
                       hep-ph/9803297;\\
                       A.~Pilaftsis,
                       {\em Phys. Lett.} {\bf B 435} (1998) 88,
                       hep-ph/9805373.

\bibitem{mhiggsCPXEP1} D.~Demir, 
                       {\em Phys. Rev.} {\bf D 60} (1999) 055006,
                       hep-ph/9901389.

\bibitem{mhiggsCPXEP2} S.~Choi, M.~Drees and J.~Lee,
                       {\em Phys. Lett.} {\bf B 481} (2000) 57,
                       hep-ph/0002287.

\bibitem{schwachnath} T.~Ibrahim and P.~Nath,
                      {\em Phys. Rev.} {\bf D 63} (2001) 035009, 
                      hep-ph/0008237;
                      {\em Phys. Rev.} {\bf D 66} (2002) 015005,
                      hep-ph/0204092.

\bibitem{mhiggsCPXRG1} A.~Pilaftsis and C.~Wagner, 
                       {\em Nucl. Phys.} {\bf B 553} (1999) 3,
                       hep-ph/9902371.

\bibitem{mhiggsCPXRG2} M.~Carena, J.~Ellis, A.~Pilaftsis and C.~Wagner,
                       {\em Nucl. Phys.} {\bf B 586} (2000) 92,
                       hep-ph/0003180.

\bibitem{mhcMSSM} S. Heinemeyer,
                  {\em Eur. Phys. Jour.} {\bf C 22} (2001) 521,
                  hep-ph/0108059.
 
\bibitem{mhcMSSMlong} M.~Frank, S.~Heinemeyer, W.~Hollik and G.~Weiglein, 
                      {\em in preparation}.

\bibitem{MFphd} M. Frank, 
                Dissertation, Karlsruhe, Germany, 2002, see:\\
                {\tt www-itp.physik.uni-karlsruhe.de/prep/phd/} .

\bibitem{hhg} J.~Gunion, H.~Haber, G.~Kane and S.~Dawson, 
              {\em The Higgs Hunter's Guide}, Addison-Wesley, 1990.

\bibitem{hff} S.~Heinemeyer, W.~Hollik, and G.~Weiglein, 
              {\em Eur. Phys. J.} {\bf C 16} (2000) 139, 
              hep-ph/0003022.

\bibitem{feynarts} J.~K\"ublbeck, M.~B\"ohm and A.~Denner, 
                   {\em Comp. Phys. Comm.} {\bf 60} (1990) 165;\\
                   T.~Hahn, 
                   {\em Nucl. Phys. Proc. Suppl.}  {\bf 89} (2000) 231,
                   hep-ph/0005029;
                   hep-ph/0012260.\\
                   The program and the user's guide 
                   is available via {\tt www.feynarts.de} .

\bibitem{modelfile} T.~Hahn and C.~Schappacher,
                    {\em Comput. Phys. Commun.} {\bf 143} (2002) 54,
                    hep-ph/0105349.

\bibitem{formcalc}   T.~Hahn and M.~P\'erez-Victoria,
                     {\em Comput. Phys. Commun.} {\bf 118} (1999) 153,   
                     hep-ph/9807565.

\bibitem{feynhiggs} S.~Heinemeyer, W.~Hollik and G.~Weiglein, 
                    {\em Comp. Phys. Comm.} {\bf 124} 2000 76,
                    hep-ph/9812320; 
                    hep-ph/0002213.\\
                    The codes are accessible via
                    {\tt www.feynhiggs.de} .

\bibitem{pdg} Part. Data Group,
              {\em Phys. Rev.} {\bf D 66} (2002) 010001.

\bibitem{cpx} M.~Carena, J.~Ellis, A.~Pilaftsis and C.~Wagner,
              {\em Phys. Lett.} {\bf B 495} (2000) 155,
              hep-ph/0009212.

\bibitem{plehnix} V.~Barger, T.~Falk, T.~Han, J.~Jiang, T.~Li and T.~Plehn,
                  {\em Phys. Rev.} {\bf D 64} (2001) 056007, 
                  hep-ph/0101106.

\end{thebibliography}
\end{document}